\documentclass[a4paper,11pt]{article}
\usepackage{pos}

\title{CMS tracking performance in Run 2 and early Run 3}

\author*[a,b]{Walaa Elmetenawee}
\onbehalf{on behalf of the CMS collaboration}
\affiliation[a]{INFN - Sezione di Bari, Via Giovanni Amendola, 173, 70126 Bari, Italy}
\affiliation[b]{Physics Department Faculty of Science, Helwan University, Ain Helwan 11795 Cairo, Egypt}

\emailAdd{walaa.elmetenawee@ba.infn.it}

\abstract{A precise and efficient tracking is one of the critical components of the CMS physics program as it impacts the ability to reconstruct the physics objects needed to understand proton-proton collisions at the LHC. The CMS detector has undergone extensive improvements in preparation for Run 3 of the LHC to operate efficiently at the increased luminosity and pileup. Significant algorithmic enhancements have been implemented to enhance the performance of the CMS tracking system. These enhancements concentrate on refining both track finding and selection processes. Performance measurements of the track reconstruction both in simulation and collision data will be presented. The performance is assessed using LHC Run 2 at $\sqrt{s}$ = 13 TeV and early LHC Run 3 data at $\sqrt{s}$ = 13.6 TeV.}

\FullConference{The 32nd International Workshop on Vertex Detectors (VERTEX2023)\\
 16-20 October 2023\\
Sestri Levante, Genova, Italy\\}

\begin{document}
\maketitle

\section{Introduction}

The iterative tracking algorithm employed in the CMS track reconstruction process, as referenced in~\cite{sec1a} (referred to as "Iterative Tracking"), is based on the Combinatorial Kalman Filter (CKF)~\cite{sec1b}. Initial iterations focus on identifying tracks that are relatively easier to detect, such as prompt tracks with higher transverse momentum (p$_{T}$). Hits associated with these reconstructed tracks are subsequently excluded from the hit set, effectively mitigating combinatorial complexity. This strategic removal of hits facilitates the reconstruction of tracks in more challenging kinematic regions.

The track reconstruction can be described as a 4-step procedure. First, the algorithm constructs seeds. This step known as "seeding" provides the initial trajectory parameter estimates. These proto-candidates are progressed through the entire tracker, identifying compatible hits at each layer using the CKF algorithm and updating the track candidate and its parameters, a phase referred to as "building." Following the combination of all associated hits, the tracks undergo a fitting process and are annotated with quality flags during the "fitting" and "selection" steps, respectively. Figure 1 illustrates the expected tracking efficiency based on Monte Carlo (MC) studies, employing events from $t\bar{t}$ production with an average pileup of 35, as detailed in~\cite{sec1c}.

\begin{figure}[ht]
\centering
\includegraphics[width=5.5cm]{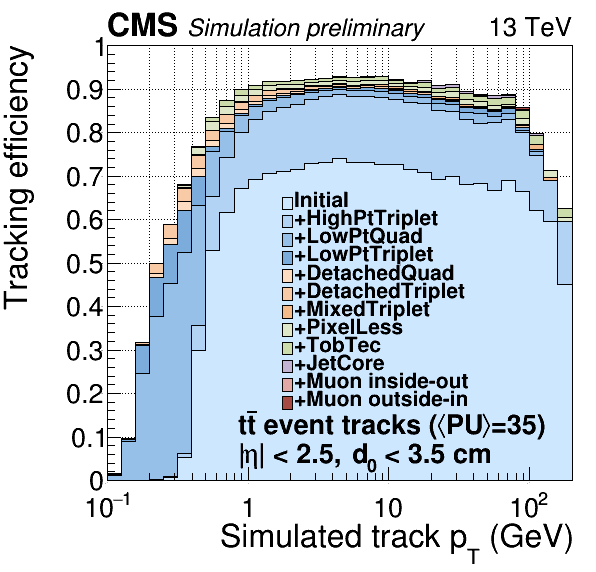}
\qquad
\caption{Tracking efficiency in MC as a function of track $p_{T}$.~\cite{sec1c}}
\end{figure}

\section{Algorithmic improvements for Run 3}
\label{improv}
In preparation for the LHC Run 3, the CMS detector has undergone extensive improvements to operate efficiently at the increased luminosity and pileup. Notable upgrades include the replacement of the innermost barrel pixel layer with new modules featuring enhanced front-end readout chips. Furthermore, significant algorithmic enhancements have been implemented to enhance the performance of the CMS tracking algorithm. These enhancements concentrate on refining both tracking finding and selection procedures. Additionally, efforts have been directed towards the development of High-Level Trigger (HLT) software, leveraging the utilization of
heterogeneous computing architectures in order to improve the
performance of the HLT both in terms of throughput and reconstructed object performance.

The track finding step experienced a significant development between the end of the LHC Run 2 and the beginning of Run 3. The CKF described above was the standard tracking algorithm employed by CMS in Run 1 and Run 2. In Run 3, a novel algorithm has been developed for track pattern recognition, known as mkFit~\cite{Sec2_b1}. This algorithm maximizes parallelization and vectorization in multi-core CPU architectures. Utilized within the CMS software for a specific subset of six tracking iterations in the referenced study~\cite{Sec2_b2}, and currently employed in the ongoing Run 3 data-taking for five iterations, the mkFit algorithm maintains physics performance comparable to traditional CKF-based pattern recognition while enhancing the computational efficiency of CMS track reconstruction.

The tracking performance is evaluated in a simulated ttbar sample with superimposed pileup (PU) events by associating reconstructed tracks and simulated tracks. The number of generated PU events follows a uniform distribution ranging from 55 to 75. A reconstructed track is considered associated with a simulated particle if more than 75\% of its hits originate from that particle~\cite{sec1a}. If this criterion is not met, the reconstructed track is classified as a random hit combination and designated as a fake track. Simulations consider detector conditions without module failures, taking into account the residual radiation damage from Run 2 operations.

Simulated tracks coming from the signal (hard scattering) vertex with a transverse momentum (p$_{T}$) greater than 0.9 GeV and pseudorapidity ($|\eta|$) less than 3.0 are used in the performance computation. Tracking efficiency is defined as the fraction of simulated tracks associated with at least one reconstructed track. All reconstructed tracks coming from any vertex (including pileup vertices) are used in the fake rate computation. The tracking fake rate is defined as the fraction of reconstructed tracks not associated to any simulated particle.

\begin{figure}[ht]
\centering
\includegraphics[width=5cm]{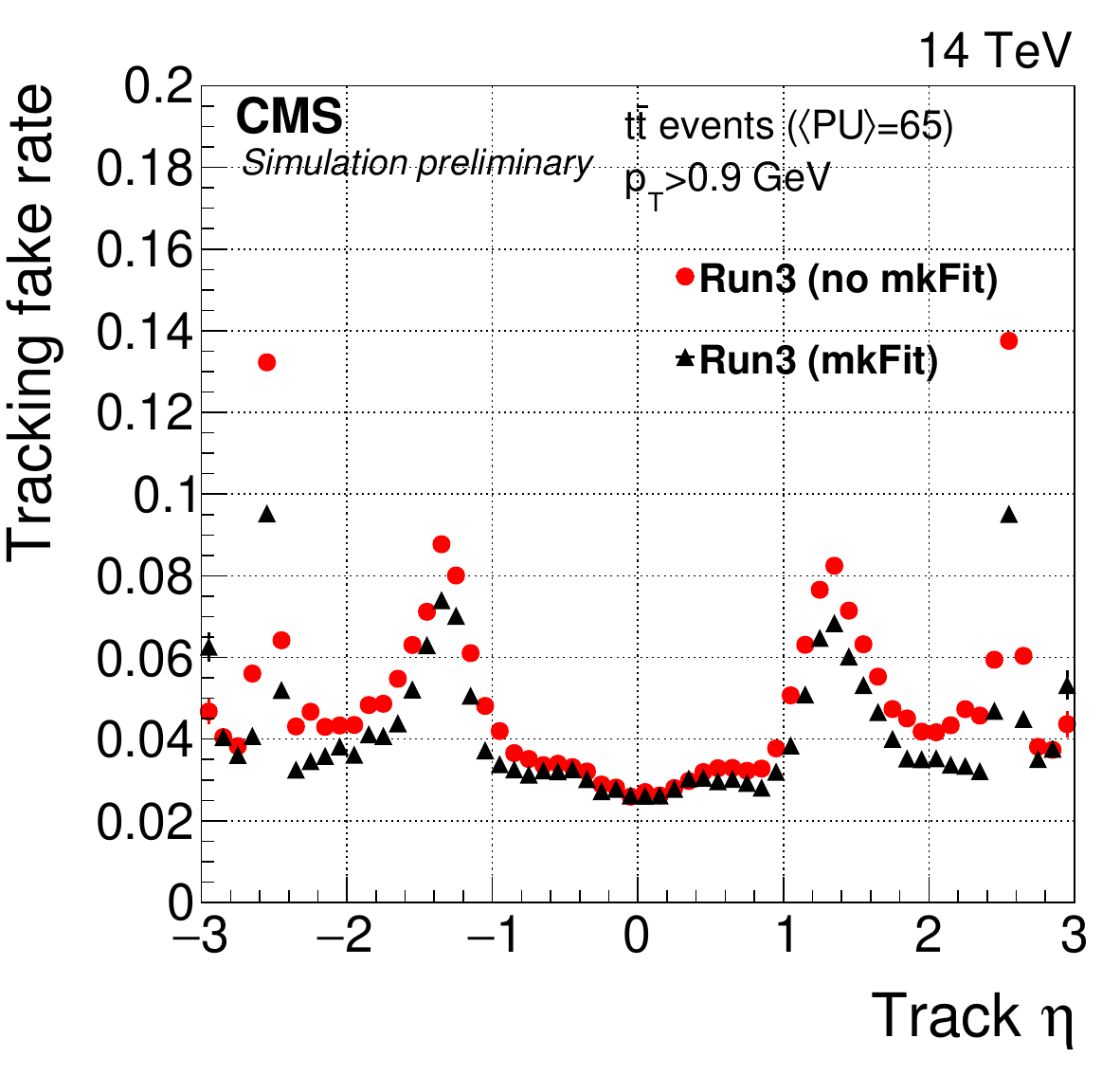}
\qquad
\includegraphics[width=5cm]{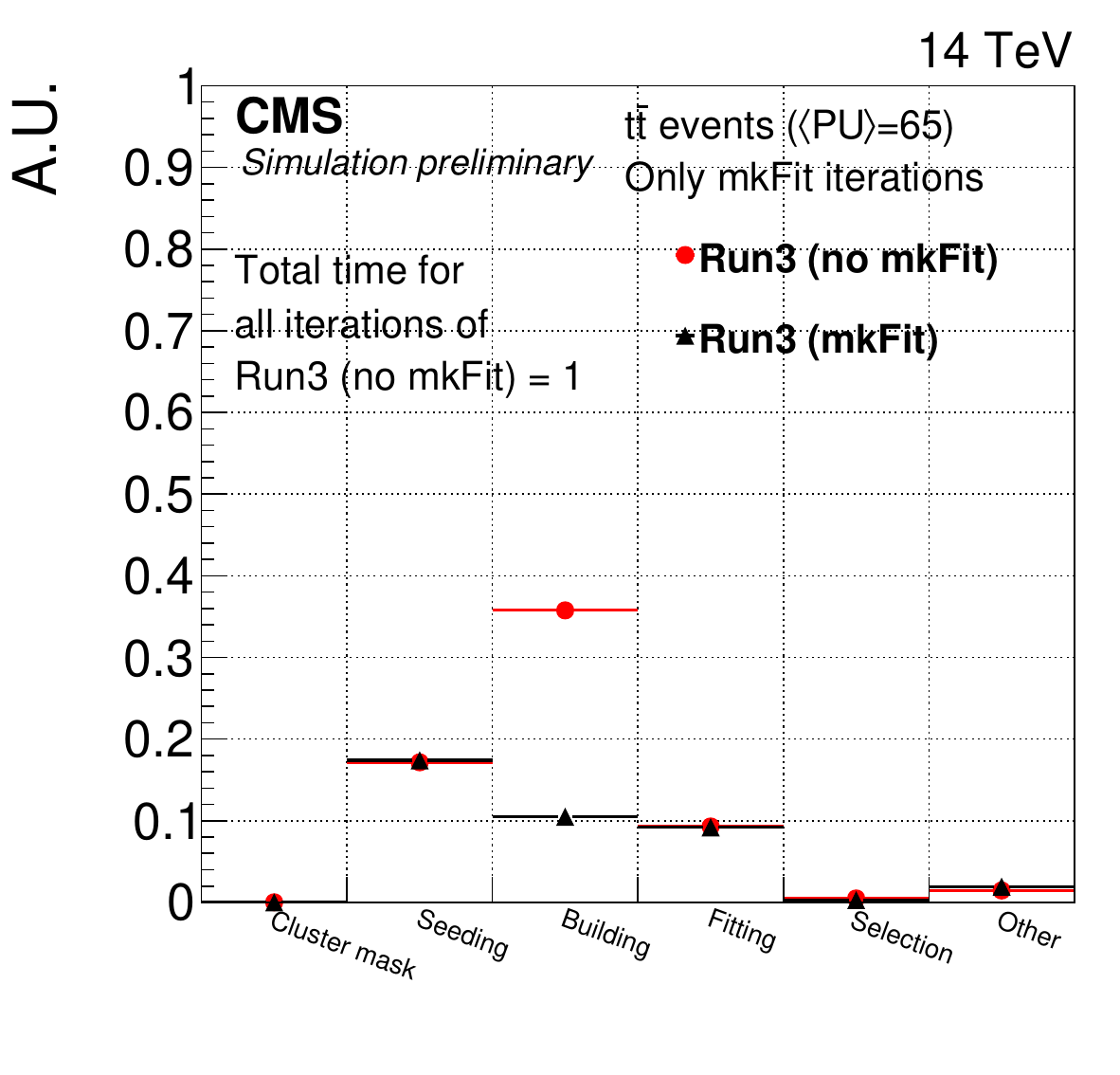}
\caption{Tracking fake rate as a function of the track pseudorapidity $\eta$ (left) and the tracking time as a function of the tracking steps, for the subset of six tracking iterations using mkFit for track building (right).~\cite{Sec2_b2}}
\label{figmkfit}
\end{figure}

The average tracking fake rate is consistently lower when utilizing mkFit for track building in a subset of six tracking iterations, as illustrated in the left panel of Figure~\ref{figmkfit}. Furthermore, employing mkFit leads to a substantial reduction in track-building time by approximately 3.5 times when considering the sum of time across these six iterations as highlighted in the right panel of Figure~\ref{figmkfit}. Examining individual iterations where mkFit is applied, the reduction factor varies, ranging from about 2.7 to approximately 6.7.

The track classification comes into play at the end of each iteration in the iterative tracking procedures for the track reconstruction process. A classifier using a multivariate analysis is applied after each iteration and several selection criteria are defined. If a track meets the high purity requirement, its hits are removed from the hit collection, thus simplifying the later iterations, and making the track classification an integral part of the reconstruction process. Tracks passing loose selections are also saved for physics analysis usage. The CMS experiment improved the track classification starting from a parametric selection used in Run 1, moving to a Boosted Decision Tree (BDT) in Run 2, and finally to a Deep Neural Network (DNN) in Run 3. The DNN was developed initially for pure CKF reconstruction, but later for both mkFit and CKF tracks, which can have slight differences. These differences required training the DNN twice using two different sets of input tracks while keeping all the training settings unchanged. Reference~\cite{Sec2_DNN1} provides more information about the training procedures. In the current default tracking a DNN trained on mkFit tracks is used for the mkFit iterations, while a DNN trained on the CKF tracks is used for the CKF iterations.

\begin{figure}[ht]
\centering
\includegraphics[width=5cm]{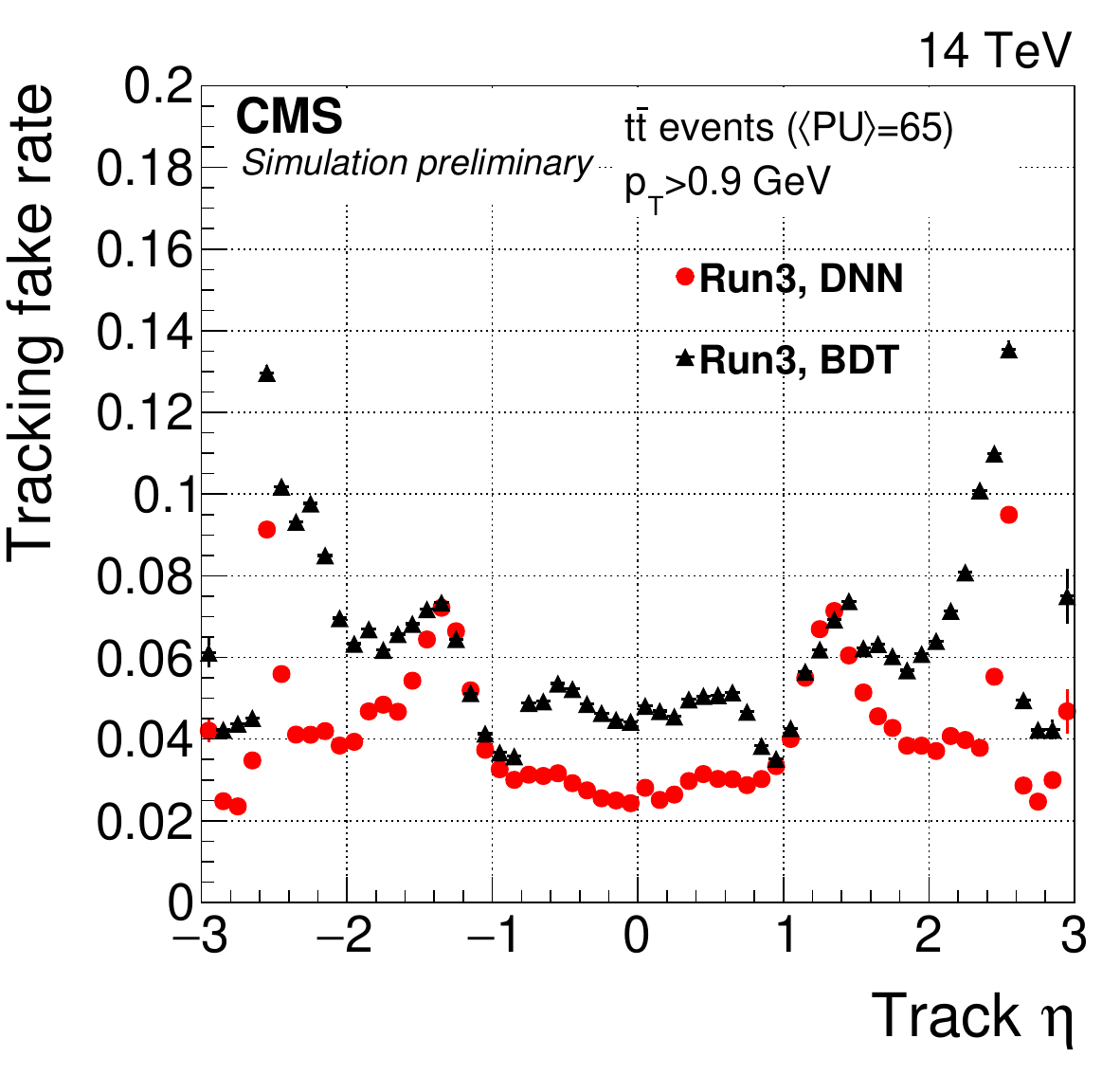}
\qquad
\includegraphics[width=5cm]{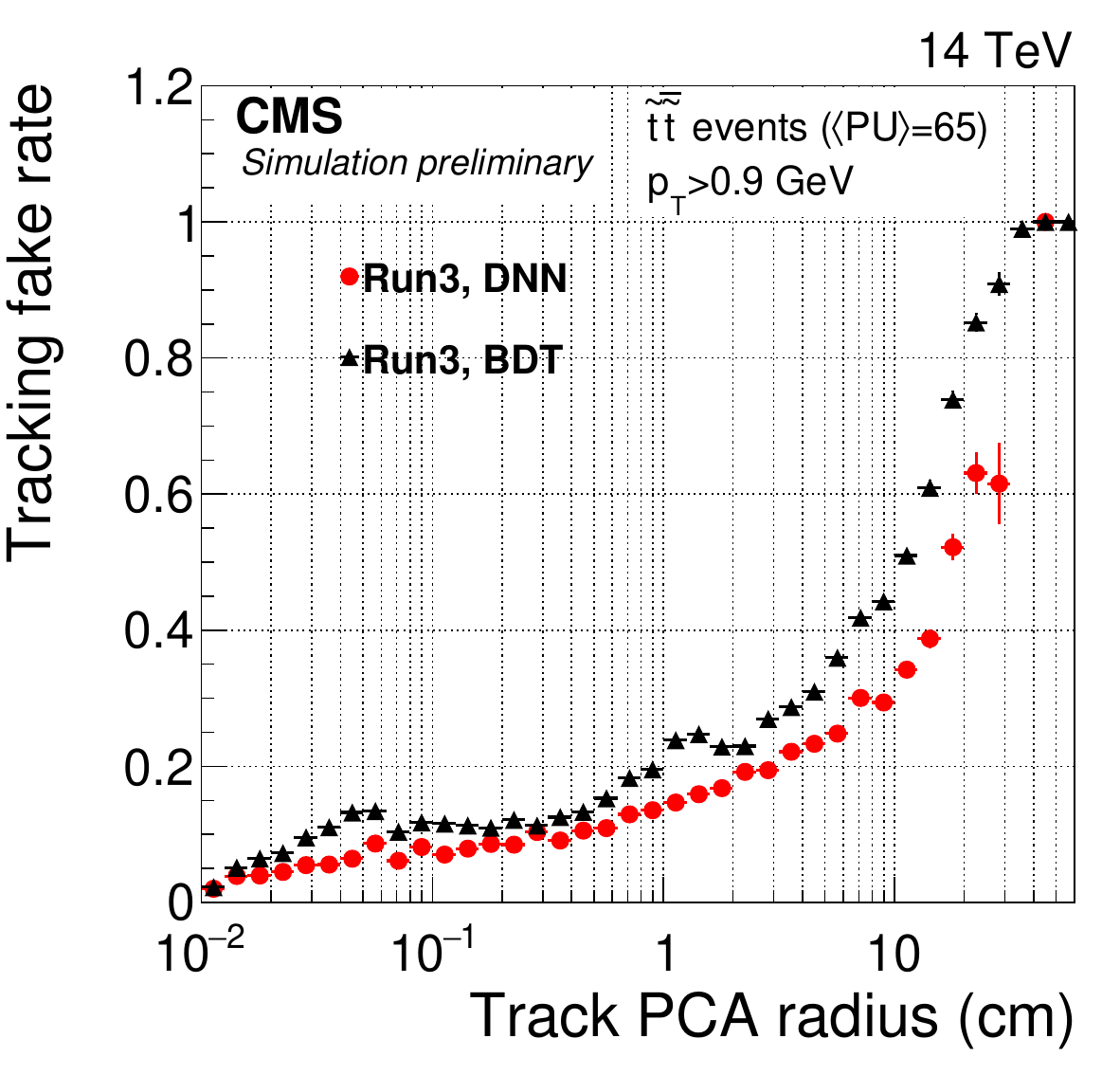}
\caption{The tracking fake rates for tracks selected by DNN (red) and BDT (black) are compared in terms of track pseudorapidity $\eta$ (left) and the radius of the track point of closest approach to the beamline (right).~\cite{Sec2_DNN1}}
\label{figDNN}
\end{figure}

The DNN performance is shown in the current default tracking (mkFit + CKF Run 3) and compared to the result of the Run 2 BDT on the same tracks. The physics results are shown after applying the high purity BDT or DNN selection to each iteration and after merging all the tracks from the iterations into one collection. The tracking fake rate when the DNN is used is lower than the one obtained using the BDT as shown in Figure~\ref{figDNN}. The largest fake rate reductions are in the tracker endcaps ($|\eta|$ > 2) and in the barrel ($|\eta|$ < 1), in addition, there is a reduction in the fake rate across all the radii values, with a reduction of about 30\%.

The HLT of the CMS employs a streamlined version of full event reconstruction designed for fast processing. In the ongoing Run 3, the HLT utilizes a heterogeneous computing farm~\cite{Sec2_c1}. During Run 3, track reconstruction at the 
HLT is based on a single iteration of the CKF algorithm~\cite{sec1b}, utilizing hits recorded by pixel and strip detectors. This single iteration is seeded by pixel tracks reconstructed through the Patatrack algorithm, which can be efficiently offloaded to Graphics Processing Units (GPUs)~\cite{Sec2_c2, Sec2_c1}. In order to be eligible as seeds, Patatrack pixel tracks must be constructed with a minimum of three pixel hits, possess a transverse momentum (p$_{T}$) exceeding 0.3 GeV, and be consistent with a leading pixel vertex.

Pixel vertices originating from primary interactions are reconstructed at the HLT during Run 3 through Patatrack pixel tracks featuring a minimum of four pixel hits and p$_{T}$ exceeding 0.5 GeV. This reconstruction process involves clustering selected pixel tracks based on their z-coordinates at the point of closest approach to the center of the reconstructed ellipsoid approximating the luminous region. Subsequently, pixel vertices are arranged in descending order according to the summed p$_{T}^{2}$ of the associated pixel tracks. The vertex with the highest summed p$_{T}^{2}$ is labeled as the primary vertex (PV).

\begin{figure}[ht]
\centering

\includegraphics[width=4.8cm]{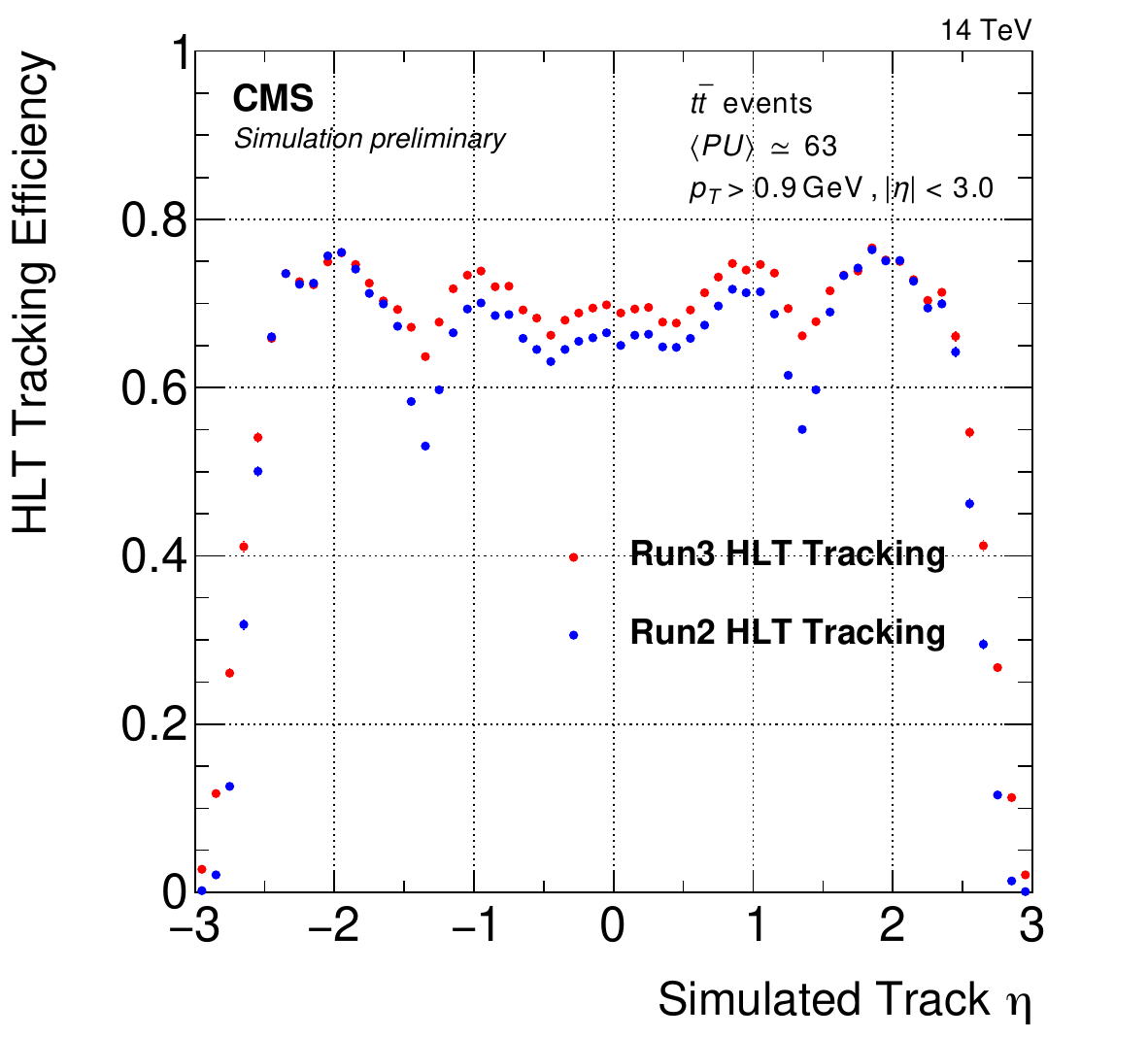}
\includegraphics[width=4.8cm]{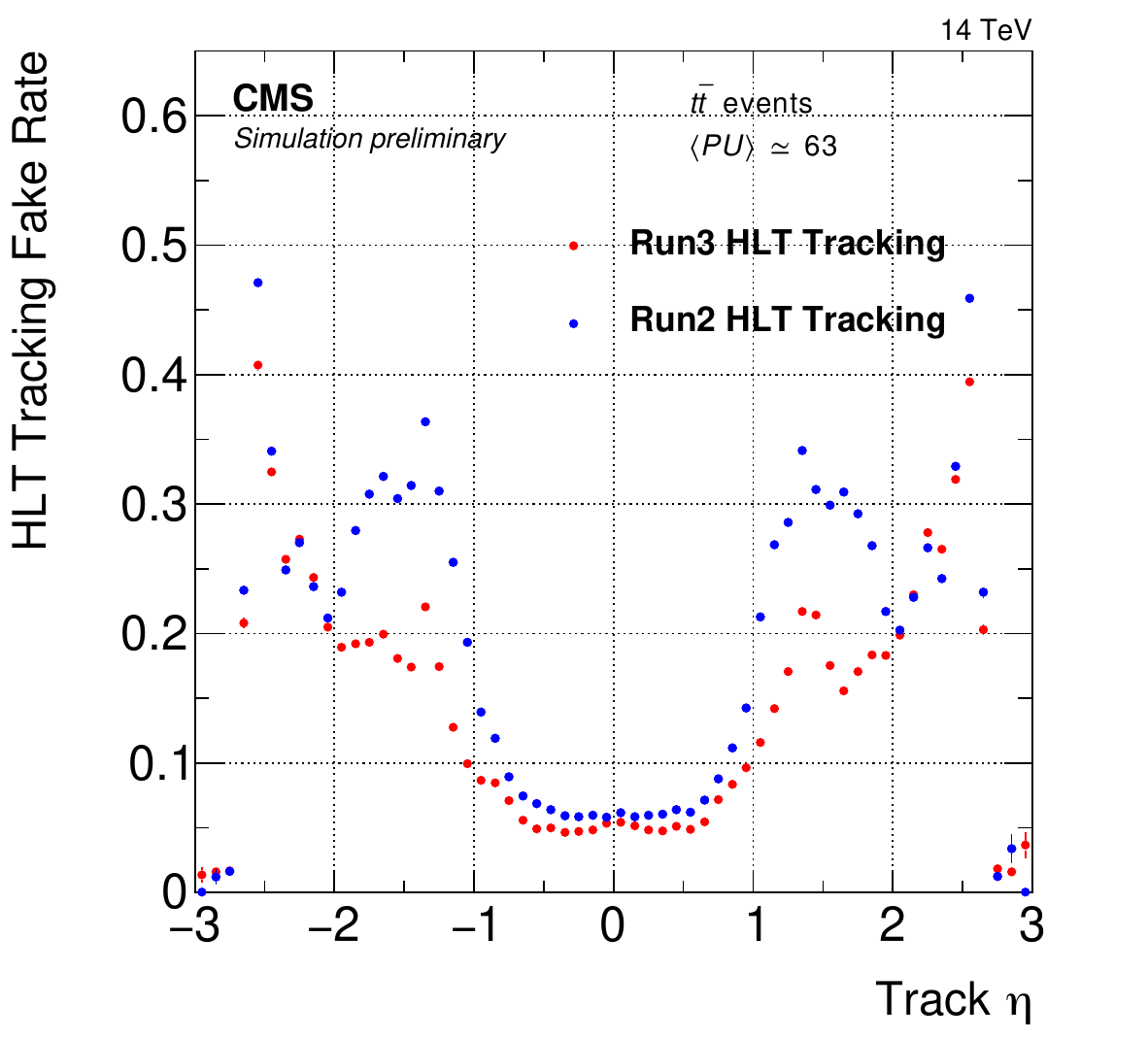}
\includegraphics[width=4.8cm]{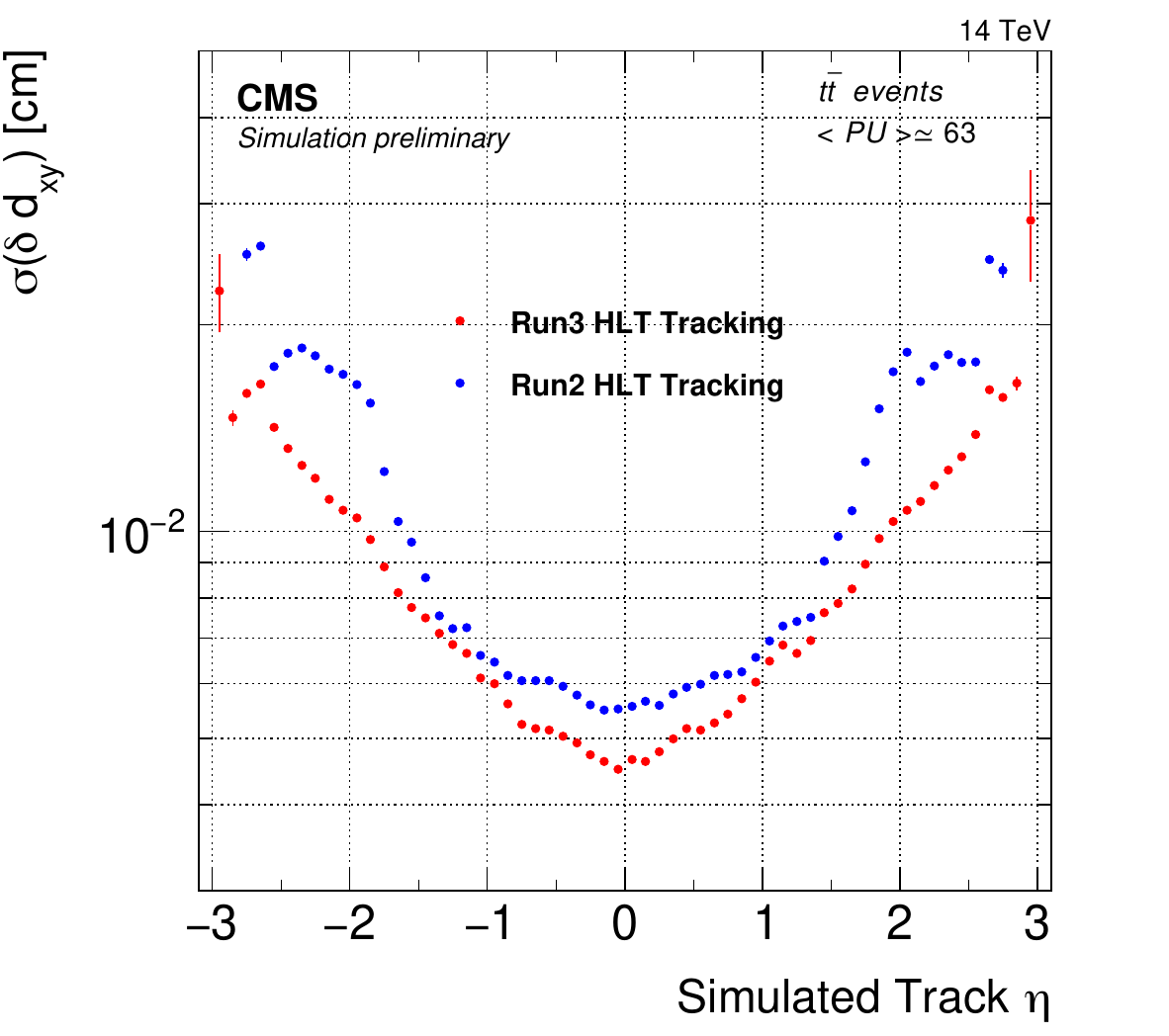}
\qquad
\caption{The plots illustrate the tracking efficiency (left), fake rate (middle), and track resolution (right) as a function of the track pseudorapidity ($\eta$) for both Run 2 HLT tracking (in blue) and Run 3 (in red).~\cite{Sec2_c2}}
\label{figSIMHLT}
\end{figure}

Figure~\ref{figSIMHLT} presents a comparative analysis of the HLT tracking performance in Run 2 and Run 3. In the left panel of Figure~\ref{figSIMHLT}, tracking efficiency is presented, where an improvement of up to 20 \% is observed in the efficiency of Run 3 compared to Run 2. Moving to the middle panel of Figure~\ref{figSIMHLT}, the tracking fake rate is showcased, demonstrating a notably smaller fake rate for Run 3 HLT tracking in the region with 0.9 < $|\eta|$ < 2.1, corresponding to the transition region between the barrel and the endcap detectors~\cite{Sec2_c2}. This is 
understood as in Run 3, pixel triplets are used to seed track reconstruction across the entire detector volume, providing improved performance in the transition regions compared to Run2 HLT tracking, in which pixel triplets were confined to seeding the third iteration in regions around jet candidates identified from calorimeter information and tracks reconstructed in the preceding two iterations. The right panel of Figure~\ref{figSIMHLT} shows the track resolution $d_{xy}$, revealing a significant improvement in Run 3 performance across all regions compared to Run 2.

\section{Tracking Performance using Tag and Probe Technique}

The tracking performance can be characterized in the data by applying the tag-and-probe technique~\cite{sec1a}. The efficiency is estimated using events with a Z boson decays into a pair of muons. The so-called "tag" muon is reconstructed using both the muon chambers and the tracker and identified by stringent requirements. The other muon ("probe"), instead, is reconstructed using only hits from the muon system and identified by looser requirements. Probes are then sorted into two categories, passing and failing, considering whether or not they can be matched to at least one track in a cone ($\Delta R$ < 0.3) around the probe muon. The efficiency is calculated as the ratio between the passing probes and the total number of probes. 

\begin{figure}[ht]
\centering

\includegraphics[width=4.8cm]{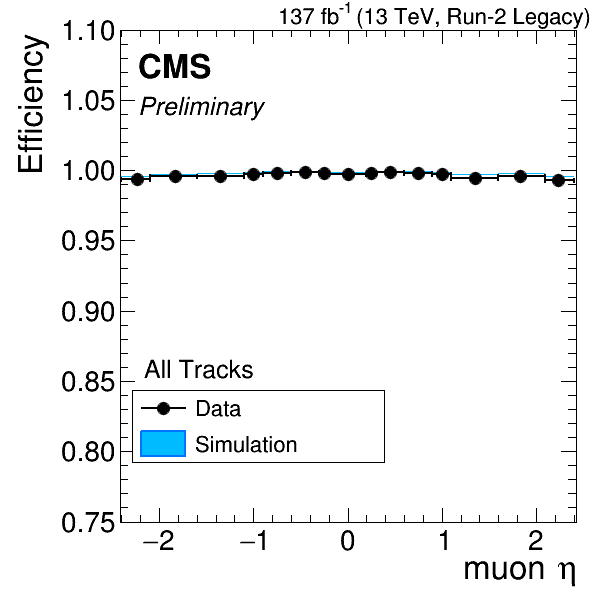}
\includegraphics[width=4.8cm]{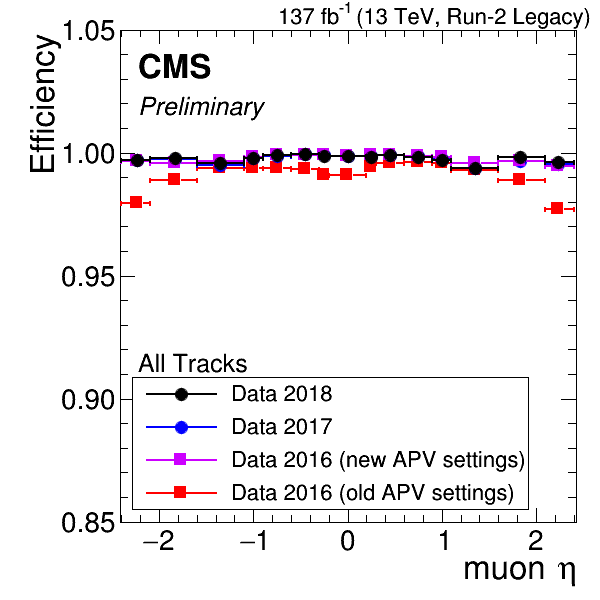}
\includegraphics[width=4.8cm]{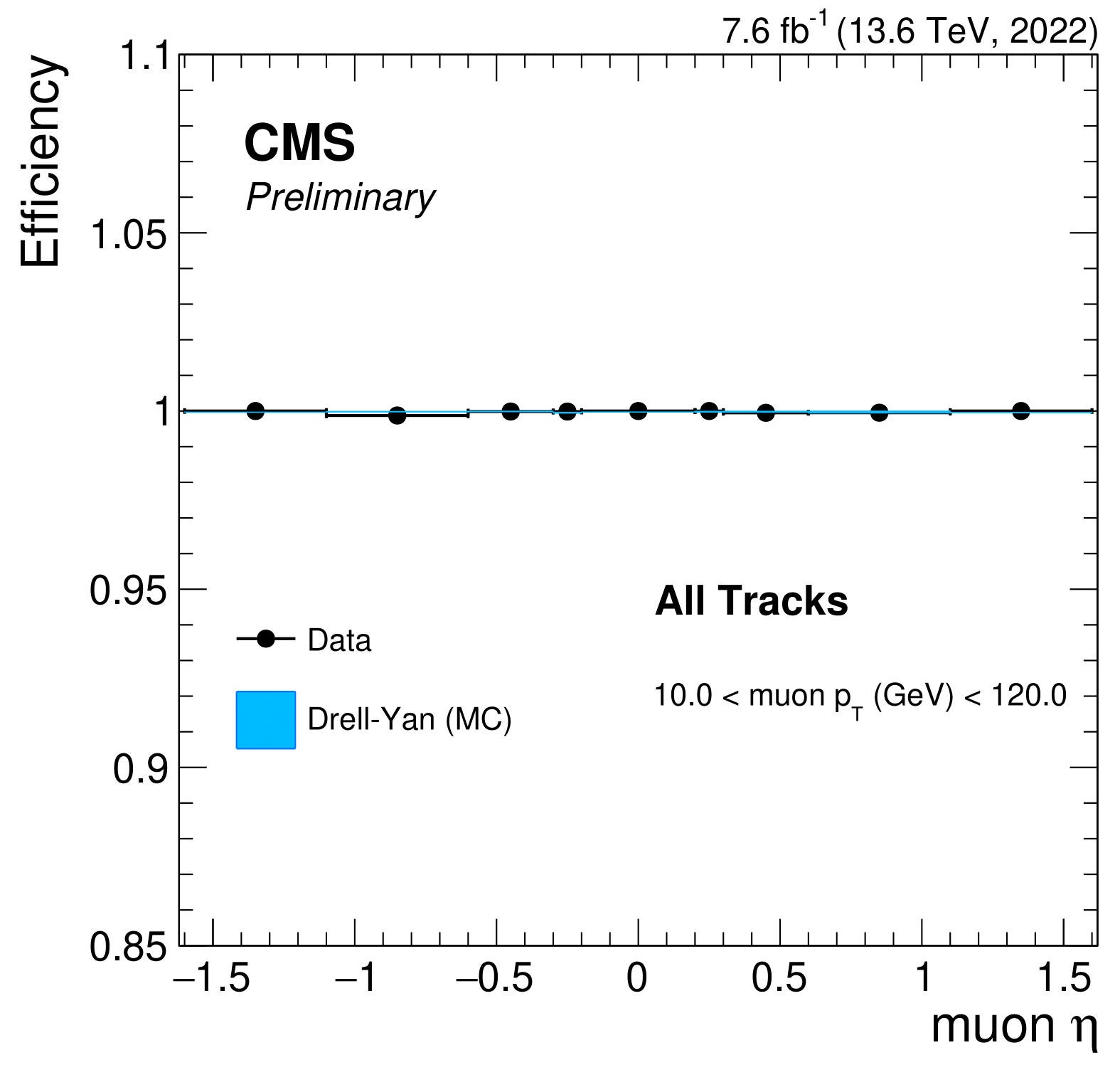}
\qquad
\caption{The tracking efficiency as a function of pseudorapidity $\eta$ of the probe muon is shown for the full Run 2 Legacy~\cite{Sec3_1} (left), the different eras of Run 2 data (middle), and for early Run 3~\cite{Sec3_2} (right).}
\label{figTnP}
\end{figure}

A tracking efficiency of about 99.9 $\%$ for tracks associated with muons has been observed in the whole muon pseudorapidity acceptance through the full Run 2 data taking, as shown in Figure~\ref{figTnP} (left). Figure~\ref{figTnP} (middle) provides a comparative analysis of tracking efficiency across different running periods of Run 2. Notably, a period in 2016, denoted "2016 old APV," exhibited an inefficiency of up to $\sim$ 2\%. This inefficiency was attributed to strip readout electronic saturation during high-intensity data-taking, a condition rectified in the remaining 2016 data, denoted as "new APV settings," following APV setting adjustments for fast recovery~\cite{Sec3_0}, resulting in a good agreement in efficiency with other data-taking periods (2017 and 2018 data). 

Thanks to algorithmic improvements detailed in the previous section and the replacement of the innermost barrel pixel layer with new modules featuring enhanced front-end readout chips to address issues identified in Run 2, the same excellent efficiency has been maintained for Run 3, as depicted in Figure~\ref{figTnP} (right) for the barrel region.

\section{Performance of tracking at High-level trigger (HLT)}

The tracking efficiency and fake rate of the HLT, as measured in experimental data, are defined w.r.t. offline tracks generated through complete offline event reconstruction, which satisfy high-purity track quality criteria~\cite{sec1a}. Both HLT and offline tracks are required to have a transverse momentum ($p_{T}$) exceeding 0.9 GeV, a transverse impact parameter relative to the primary vertex (defined in Section~\ref{improv}) of $|d_{xy}|$ < 2.5 cm, and a longitudinal impact parameter relative to the primary vertex of $|d_{z}|$ < 0.1 cm. The matching between HLT and offline tracks is established by enforcing an angular distance criterion, requiring $\Delta R = \sqrt{ (\Delta \eta)^{2} + (\Delta \phi)^{2}}$ < 0.002.

\begin{figure}[ht]
\centering
\includegraphics[width=5cm]{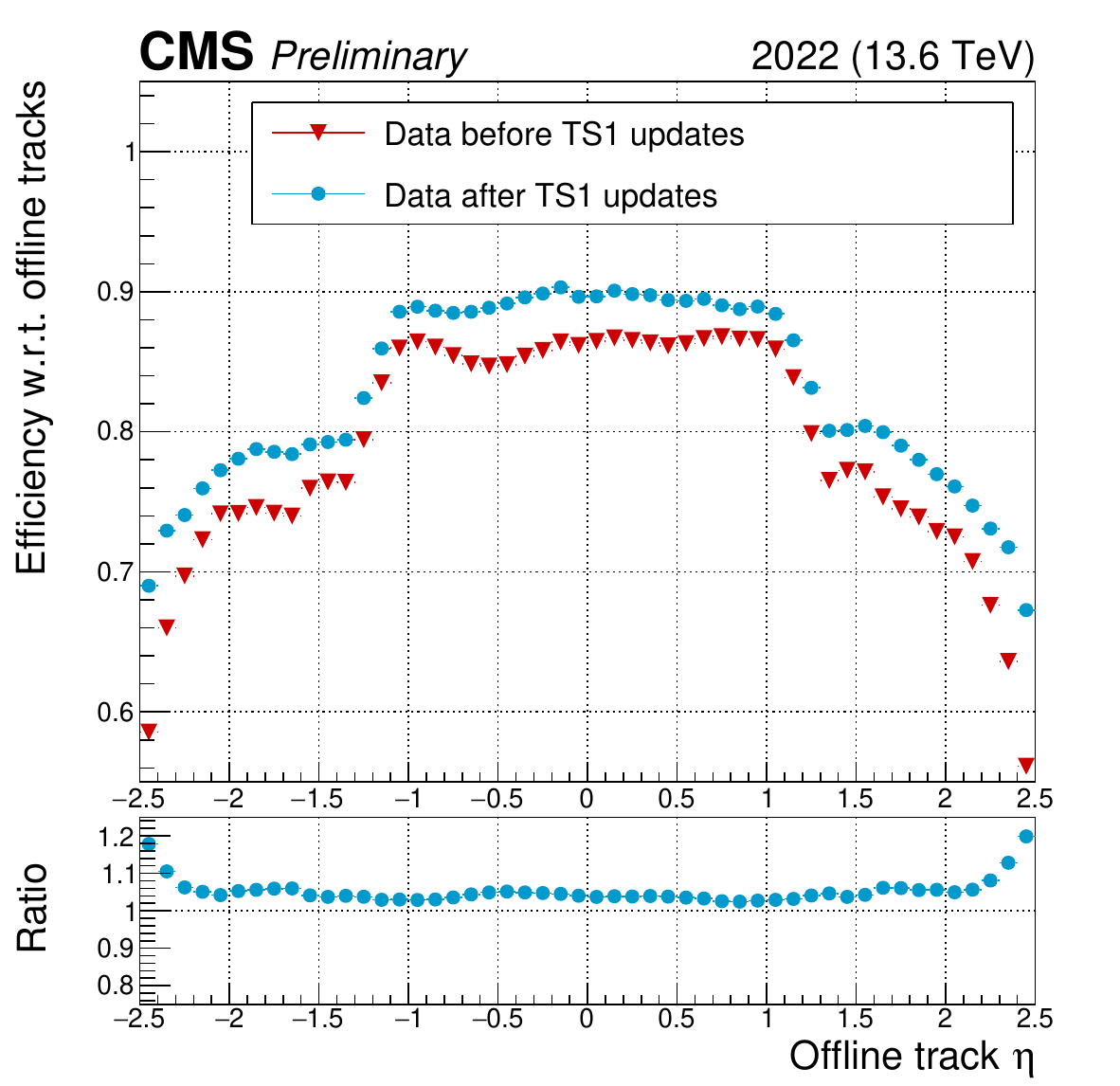}
\qquad
\includegraphics[width=5cm]{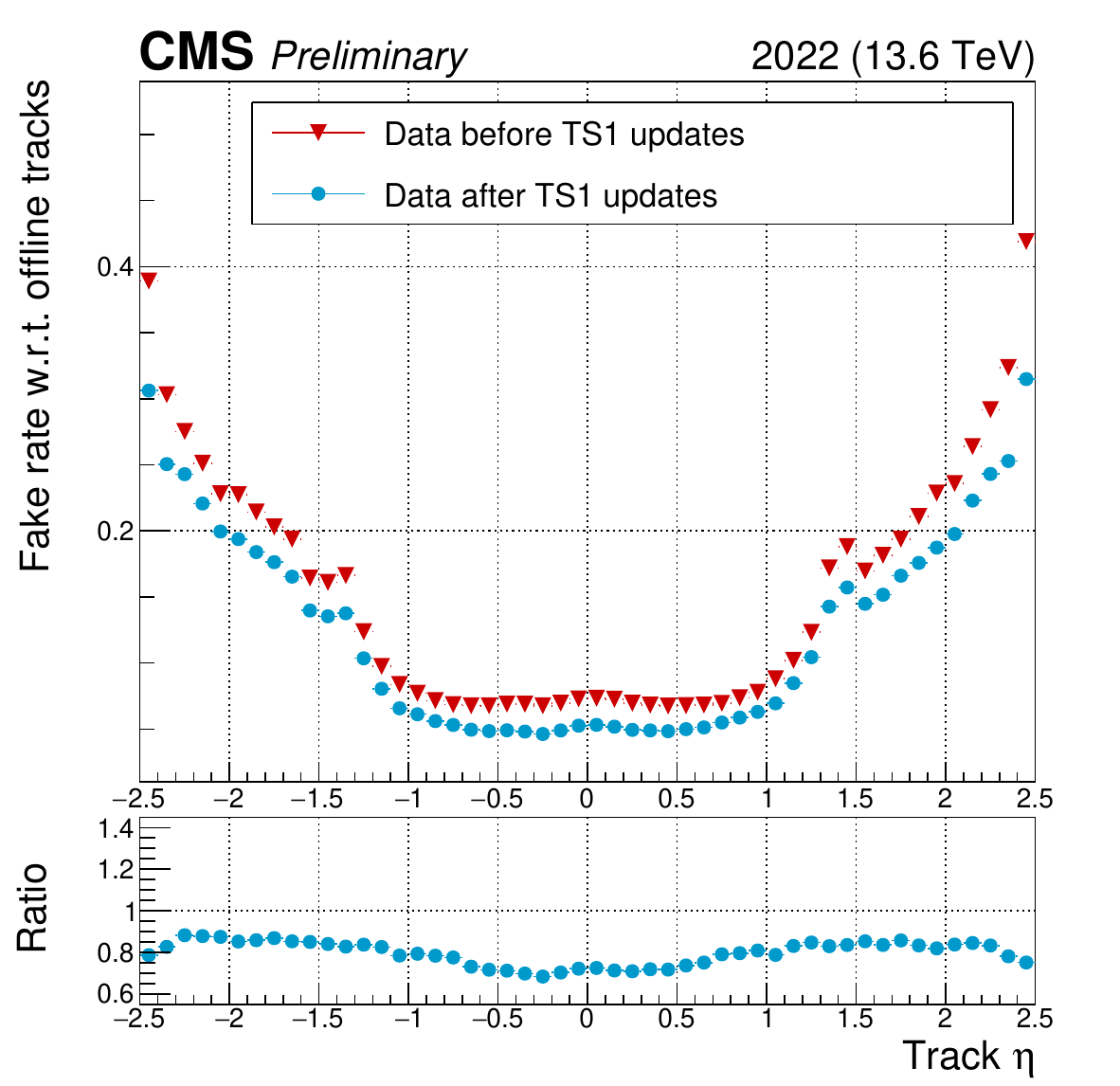}
\caption{The HLT tracking efficiency (left) and 
fake rate (right) w.r.t. offline reconstruction is shown as a function of the offline track pseudorapidity $\eta$.~\cite{Sec4_b}.}
\label{figHLT}
\end{figure}

The HLT tracking efficiency w.r.t. offline tracks is defined as the ratio between the number of HLT tracks matched to an offline track according to the requirement described above and the total number of offline tracks, while the fake rate is the ratio of HLT tracks that are not associated to any offline track and the total number of HLT tracks. Tracking efficiency (fake rate) relative to offline tracks, measured in data collected shortly before and after the LHC Tecnical Stop 1 (TS1), is represented as a function of the pseudorapidity of the offline track in Figure~\ref{figHLT} left (right). Discrepancies in efficiency across the entire range arise from variations in efficiency in BPix L1.

\section{Conclusions}
Significant algorithmic enhancements in the track finding and selection have been implemented to enhance the performance of the CMS tracking in preparation for the LHC Run 3, resulting in reduced tracking fake rates and total tracking CPU time compared to Run 2. The mentioned improvements together with the installation of the 
replaced innermost pixel detector layer, effectively address the efficiency challenges, ensuring the CMS tracking system excels in the ongoing Run 3.

\end{document}